\documentstyle[prd,aps,preprint,epsfig,amsmath]{revtex}
\begin{document}
\tighten
\voffset=1cm
\preprint{UMD PP\#01-145}
\preprint{DOE/ER/40762-229}
\title{Photon Structure Functions from Quenched Lattice QCD}
\author{Xiangdong Ji and Chulwoo Jung}
\address{Department of Physics,
University of Maryland,
College Park, Maryland 20742 }
\date{\today}          
\maketitle
\begin{abstract}
We calculate the first moment of the photon structure function, 
$\langle x\rangle^{\gamma}=\int^1_0 dx F^{\gamma}_2(x,Q^2)$, 
on the quenched lattices with $\beta=6.0$ using the 
formalism developed by the authors recently. In this exploratory 
study, we take into account only the connected contractions. 
The result is compared with the experimental data as well 
as model predictions.
\vspace{1cm}
\end{abstract}


The parton (quark and gluon) distributions in hadrons, which can 
be measured from lepton-hadron deep inelastic scattering and other 
hard processes, have played a crucial role in 
understanding high-energy scattering and the hadron structure.
Recently it has become possible to calculate moments of 
these distributions from first principles
using lattice QCD \cite{martinelli,schierholz,negele}.

There are many other physical observables which are yet to be
calculated from lattice QCD. A special class of 
of these involve either virtual or real photons. 
Until recently, the conditions under which these 
quantities can be calculated using
lattice QCD remained unclear. 
The reason is that the photon is not an eigenstate of QCD.
Rather, a ``photon'' state
in nature is a superposition of the U(1) gauge boson 
and quark-gluon configurations which are suppressed 
by the electromagnetic coupling. 
The matrix elements of QCD operators in the photon states 
are time-dependent correlations which are defined in 
the Minkowski space, and the standard method 
used for calculating hadron matrix elements 
on the lattice is not directly applicable 
\cite{martinelli,schierholz,negele}.

In our earlier paper \cite{photon00}, we showed that the matrix element of
a quark-gluon operator between photon states can be evaluated using
lattice QCD. The relevant expression is
\begin{equation}
    \langle \gamma(p\lambda')|{\cal O}(0)| \gamma(p\lambda)\rangle 
= -e^2 \int d^4x d^4y
         e^{\omega (x_4-y_4)} e^{-i\vec{p} \cdot (\vec{x}-\vec{y})}
       \langle 0 |T_E \epsilon^{*}(\lambda')\cdot J(x) {\cal O}(0) 
\epsilon(\lambda) \cdot J(y)|0\rangle 
\label{start_exp}
\end{equation}
where every quantity has been expressed in the Euclidean 
space: $T_E$ is the Euclidean time-ordering; 
the electromagnetic current $J_\mu (\mu=1,2,3,4)$ is defined as 
$\sum_f e_f \overline\psi_f\gamma_\mu\psi_f$ with Euclidean $\gamma$ 
matrices $\gamma_i = -i\gamma_M^i (i=1,2,3)$ and 
$\gamma_4 = \gamma^0$; the Euclidean photon 
polarization vector is defined as $\epsilon_i 
= \epsilon^i_M$ and $\epsilon_4 = 
i\epsilon^0 (\epsilon^{*4}=i\epsilon^{*0})$; 
$i\omega$ is the Euclidean-space photon energy.
[In Ref.~\cite{photon00}, we did not continue the electromagnetic
current and the polarization vector to Euclidean space and 
superficially the result there differs by an overall sign.]
The above expression is the same as the naive analytic 
continuation of the matrix element from the 
Minkowski space. However, Eq. (\ref{start_exp}) is 
only valid when there exists an energy gap between the 
photon energy and the lowest hadronic state 
of the same quantum number. Otherwise, the integral
becomes divergent. 

One of the quantities we can study using the above formalism 
is the photon structure functions, which can be measured 
from collisions between real or virtual photons with 
highly virtual ones, achievable in $e^+e^-\rightarrow 
e^+e^-+$hadrons~\cite{brodsky,nisius}.
While a photon is normally considered a structureless particle, 
it can fluctuate into a charged fermion-antifermion pair
or more complicated hadronic states 
which can be revealed through interactions with a highly
virtual photon. In fact, the photon structure functions 
can be defined from its hadronic tensor 
and the moments of the structure 
functions can be obtained from operator product expansion,
much in the same way as for the hadron structure functions.
There exists a substantial amount of experimental data 
for unpolarized structure functions already \cite{nisius};
and future experiments from HERA and $e^+ e^-$ are 
expected to measure the polarized structure 
functions as well \cite{polpho}.
(For a recent review on theoretical and experimental 
progress on this subject, refer to Ref.~\cite{nisius}.)

Here we report on the first lattice study of the 
structure function $F^{\gamma}_2$. 
More specifically, we measure the first moment of 
$F_2^{\gamma}$ which 
corresponds to the average fractional momentum 
carried by the quarks 
(weighed with the fractional charge squared for each flavor) in a real photon. The measurement is done on $\beta=6.0$ lattices, 
corresponding to $Q^2 \sim a^{-2} \sim (2.4~\mbox{GeV})^2$, 
in the quenched approximation for three degenerate 
flavors of quarks. Also, we choose to ignore the contributions 
from disconnected diagrams in which the quark lines do not 
connect any of the external vertices. 

We begin by briefly reviewing the definition of the photon 
structure functions and their relation to the cross sections. 
We closely follow \cite{peterson80} for the definition of 
$F^\gamma_2(x, Q^2)$. For the deep inelastic $e + 
\gamma$ process in a frame where the incoming electron
and the photon are collinear, we write
\begin{eqnarray}
e^{\pm}(k) + \gamma(p) \rightarrow e^{\pm}(k') + X(p_X) \ , 
\nonumber \\
k^\mu = (E,0,0,E) \ , \\
{k'}^{\mu} = (E',0,E'\sin\theta,E'\cos\theta)  \ ,\nonumber \\
p^\mu = (E_{\gamma},0,0,-E_{\gamma}) \ . \nonumber 
\end{eqnarray}
The relevant kinetic variables are $Q^2= -(k-k')^2$, 
$\nu= p\cdot q$, $q=k-k'$, $x=Q^2/2(p\cdot q)$, 
$y = q \cdot p /k \cdot p$.
The photon structural information is contained in the following
tensor,  
\begin{eqnarray}
W^{\mu\nu}
& =& \frac{1}{4\pi} \cdot \frac12\sum_{X,\lambda} \langle \gamma(p\lambda)|J^{\mu}(0)|X\rangle\langle X|J^{\nu}(0)|\gamma(p\lambda)\rangle (2\pi)^4\delta^4(p_X-p-q) \nonumber \\
& =& \frac{1}{4\pi} \cdot \frac12 \sum_{\lambda} \int \langle\gamma(p\lambda)|[J^{\mu}(x),J^{\nu}(0)]|\gamma(p\lambda)\rangle e^{iqx} d^4x \\ \nonumber
&=& W^\gamma_1(\nu,Q^2)\left[-g^{\mu\nu} + \frac{q^{\mu}q^{\nu}}{q^2} \right]
+ W^\gamma_2(\nu,Q^2)\left[p^{\mu} - \frac{p \cdot q}{q^2} 
q^{\mu}\right]\left[p^{\nu} - \frac{p \cdot q}{q^2} q^{\nu}\right].
\end{eqnarray}
Here $J_{\mu}(x) = \sum_f e_f \overline\psi_f(x) \gamma_{\mu} \psi_f(x)$ is 
the electromagnetic current and $e_f$ is the fractional charge 
of the quark of flavor $f$. The covariant normalization
has been used for the photon state,
$\langle \gamma(p\lambda)|\gamma(p'\lambda')\rangle = (2\pi)^3 
2\omega_{\vec{p}}\delta(\vec{p}-\vec{p'})\delta_{\lambda,\lambda'}$. 

Since we are interested in the 
unpolarized photon properties here, we have averaged over 
the photon polarization $\lambda$ in the above equations.
[The polarized photon structure functions are interesting by
themselves and can be studied in a similar approach \cite{polpho}.]

Using these variables, the differential cross section is given by
\begin{equation}
\frac{d\sigma}{dE'd\cos\theta} = 
\frac{8\pi\alpha^2_{\rm em} {E'}^2}{Q^4E_{\gamma}}
\bigg[2E^2_{\gamma}W^\gamma_2(\nu,Q^2) \cos^2 
\left( \frac{\theta}2\right) + W^\gamma_1(\nu,Q^2) 
\sin^2 \left(\frac{\theta}2\right) \bigg] \ , 
\end{equation}
or, in terms of $x$ and $y$ and scaling functions,
\begin{equation}
\frac{d\sigma}{dxdy} = \frac{16\pi\alpha_{\rm em}^2EE_{\gamma}}
{Q^4}\bigg[(1-y)F_2^{\gamma}(x,Q^2)+xy^2F^{\gamma}_1(x,Q^2)\bigg] \ ,
\end{equation}
where the dimensionless scaling functions are defined
as $F_2^{\gamma}(x,Q^2) = \nu W^\gamma_2(\nu, Q^2)$ 
and $F^{\gamma}_1(x,Q^2) = W^\gamma_1(\nu,Q^2)$.

In the parton model, the photon scaling function $F_2^\gamma(x)$
is related to the quark distributions $q_f^\gamma(x)$ in the 
photon---the probability of finding a quark with momentum fraction $x$, 
\begin{equation*}
F^{\gamma}_{2}(x) = x \cdot \sum_f e_f^2 (q_f^\gamma(x)+
\overline{q}_f^\gamma(x)) \ .
\end{equation*}
Taking into account QCD radiative corrections, the moments
of the photon structure functions can be written in the
following factorized form,
\begin{equation*}
\int^1_0 x^{n-2}F_2^{\gamma}(x,Q^2)\ dx = \sum_f C_{2n,f}
\left(\frac{Q^2}{\mu^2},g(\mu)\right)
\langle x^{n-1} \rangle^{\gamma}_f\ , ~~~~(n=2,4,...,) 
\end{equation*}
where $C_{2n,f}$ are the perturbative coefficient functions and
$\langle x^{n-1}\rangle^\gamma_f$ are the nonperturbative
moments of the parton
(quark and gluon) distributions ($f$ here refers to gluons
as well). For quarks, the moments can be calculated as the matrix
elements of local operators,
\begin{gather}
\frac12 \sum_{\lambda} \langle\gamma(p\lambda)| 
{\cal O}_{\{\mu_1\cdots\mu_n\},f}
 | \gamma (p\lambda) \rangle = 2\langle x^{n-1} 
\rangle^{\gamma}_f[p_{\mu_1}\cdots p_{\mu_n } - {\rm traces}]\ , \nonumber \\
{\cal O}_{\mu_1\cdots\mu_n, f} = \left(\frac{i}{2}\right)^{n-1}
\overline\psi_f\gamma_{\mu_1}\stackrel{\leftrightarrow}{D}_{\mu_2}\cdots
\stackrel{\leftrightarrow}{D}_{\mu_n } \psi_f \ , 
\label{mme}
\end{gather}
where $ \stackrel{\leftrightarrow}{D}_\mu = 
\stackrel{\rightarrow}{D}_\mu-\stackrel{\leftarrow}{D}_\mu$
is the covariant derivative and
$\{\cdots\}$ denotes the symmetrization of indices. 

In the following discussion, we ignore the 
radiative corrections to the coefficient functions and 
take $C_{2n,f}=e_f^2$
for quarks and 0 for gluons. We then define 
$\langle x \rangle^{\gamma}=\int^1_0 
F_2^\gamma(x,Q^2)dx$, which can be calculated as the matrix 
elements of the following operator;
\begin{equation}
{\cal{O}}^{\rm (cont)}_{\mu\nu}(x)= 
\sum_f e_f^2 \cdot \frac{i}2\overline\psi_f(x)\gamma_{\{\mu}
\stackrel{\leftrightarrow}{D}_{\nu\}}\psi_f(x) \ . 
\label{contoper}
\end{equation}
In lattice QCD, we need to define the 
corresponding lattice operator in Euclidean space:
\begin{gather}
{\cal{O}}^{\rm (latt)}_{\mu\nu}(x)
= \sum_{f,y} e_f^2 ~\left[
 \overline\psi_f(x)\Gamma^{\cal{O}}_{\mu\nu}(x,y)\psi_f(y)
    + \overline\psi_f(y)\Gamma^{\cal{O}}_{\mu\nu}(y,x)\psi_f(x)
      \right] \ , \nonumber \\
\Gamma^{\cal{O}}_{\mu\nu}(x,y) = \frac{i}{8}\left[
  \gamma_{\mu}(U_{\nu}(x)\delta_{y,x+\hat{\nu}} -
 U^{\dagger}_{\nu}(y)
\delta_{y,x-\hat{\nu}})+ (\mu \leftrightarrow \nu)\right] \ ,
\end{gather}
where the lattice spacing($a$) is set to 1 and the lattice covariant derivatives are 
defined as
\begin{gather*}
 \stackrel{\rightarrow}{D}_{\mu}\psi(x) =\frac1{2}(U_{\mu}(x)
\psi(x+\hat{\mu})-U^{\dagger}_{\mu}(x-\hat{\mu})\psi(x-\hat{\mu}))\ , \\
\overline \psi(x)\stackrel{\leftarrow}{D}_\mu=\frac1{2}(\overline\psi(x+\hat{\mu})
U^{\dagger}_{\mu}(x)-\overline \psi(x-\hat{\mu})U_{\mu}(x-\hat{\mu}))\ . 
\end{gather*}
The corresponding Euclidean matrix element to Eq.~(\ref{mme})
is 
\begin{equation}
       \frac12\sum_\lambda \langle\gamma(p\lambda) | 
     {\cal{O}}^{\rm (latt)}_{\mu\nu}|\gamma(p\lambda)\rangle
    = 2i\langle x\rangle^\gamma p_{\mu E} p_{\nu E} \ , 
\end{equation}
where the photon Euclidean four-momentum is defined
as $p_{\mu E} = (\vec{p}, i\omega)$. 

The photon matrix element in lattice QCD is
\begin{multline}
\langle \gamma(p\lambda')|{\cal O}_{\mu\nu}(0)| \gamma(p\lambda)\rangle^{\rm(latt)}
= -e^2\epsilon_\alpha^*(\lambda') \epsilon_\beta(\lambda) \\
 \times \sum_{x,y}
 e^{\omega (x_4-y_4)} e^{-i\vec{k} \cdot (\vec{x}-\vec{y})}
 \langle 0 |T_E J_\alpha(x) {\cal O}_{\mu\nu}^{\rm(latt)}(0) 
J_\beta(y)|0\rangle \ . 
\end{multline}
The Euclidean Green's function can be calculated as follows,
\begin{align}
\langle 0 |T_E J_\alpha(x) {\cal O}_{\mu\nu}^{\rm(latt)}(0) J_\beta(y)|0\rangle 
=  - \sum_{z,f,f',f''}&e_f e_{f'}e_{f''}^2\nonumber \\
\times \langle0|T_E\overline
\psi_f(x)\gamma_{\alpha}\psi_f(x)
[\overline\psi_{f''}(0)\Gamma^{\cal O}_{\mu\nu}(0,z) \psi_{f''}(z)&
+\overline\psi_{f''}(z)\Gamma^{\cal O}_{\mu\nu}(z,0)\psi_{f''}(0)]
\overline\psi_{f'}(y)\gamma_{\beta}\psi_{f'}(y)|0\rangle\nonumber \\
= - \sum_{z,f} e_f^4  (2\kappa)^3 \biggl[\langle{\rm Tr}[\gamma_{\alpha}S(x,0)
\Gamma^{\cal O}_{\mu\nu}(0,z)S(z&,y)\gamma_{\beta}S(y,x)] \rangle_g
\label{eq:trace}
 \\
 + \left\langle {\rm Tr}[\gamma_{\beta}S(y,0)\Gamma_{\mu\nu}^{\cal O}(0,z)S(z,x)
\gamma_{\alpha}S(x,y)] \right\rangle_g &  
+ \left\langle{\rm Tr}[\gamma_{\alpha}S(x,z)
\Gamma_{\mu\nu}^{\cal O}(z,0)S(0,y)\gamma_{\beta}S(y,x)] \right\rangle_g
\nonumber \\
 + \langle{\rm Tr}&[\gamma_{\beta}S(y,z)\Gamma_{\mu\nu}^{\cal O}(z,0)S(0,x)
\gamma_{\alpha}S(x,y)] \rangle_g \biggr]
+ ... \ ,  \nonumber 
\end{align}
where we have shown explicitly the connected contraction
with one trace and the ellipsis represents 
disconnected diagrams with more than one
trace, and $\langle\cdots\rangle_g$ denotes
averaging over gauge configurations. 
The propagator $S(x,y)$ is solved from $\sum_x D(y,x)S(x,z)=\delta_{y,z}$ where 
\begin{equation}
D(x,y)= \delta_{x,y}-\kappa\sum_{\mu}[(1-\gamma^{\mu})
U_{\mu}(x)\delta_{y,x+\hat{\mu}}+(1+\gamma^{\mu})U^{\dagger}_{\mu}
(y)\delta_{x,y+\hat{\mu}}] \ , 
\end{equation}
is the scaled Wilson-Dirac operator. 
The Dirac and color indices are suppressed for 
readability and the traces are over those indices.

In this exploratory study, we will ignore the 
disconnected contribution. For the connected contraction,
we can show, using the charge conjugation property of the 
Wilson-Dirac operator, that the average traces for the left and
the right derivatives are the same and, hence, we only need 
to evaluate one of them. The Wilson-Dirac operator 
$D(x,y)$ is invariant under the charge conjugation:
\begin{equation}
\psi(x) \rightarrow C\overline\psi(x)^T, \qquad\ 
\overline\psi(x) \rightarrow -\psi(x)^T C^{-1}, \qquad\
U_\mu(x) \rightarrow (U_\mu(x))^*\ ,
\label{eq:cc}
\end{equation}
where the matrix $C$ satisfies $C^{-1}\gamma_\mu C = -\gamma_\mu^T$ and $T$
denotes transpose operation.
Using Eq.~(\ref{eq:cc}), we can show
\begin{equation}
S(x,y|U)^T = C^{-1}S(y,x|U^*)C\ ,
\end{equation}
where we explicitly denote
the dependence of the propagator on the gauge field.

Now we can relate the traces which arise from the left derivative of
$\Gamma_{\mu\nu}^{\cal O}$ with those from the right derivative. For example,
\begin{align}
& {\rm Tr}\left[\gamma_{\alpha}S(x,0|U) \gamma_{\nu}U_\mu(0) S(\hat{\mu},y|U)
\gamma_{\beta}S(y,x|U)\right] \nonumber \\
&={\rm Tr}\left[S(y,x|U)^T\gamma_{\beta}^TS(\hat{\mu},y|U)^T U^T_\mu(0)\gamma_{\nu}^T 
S(x,0|U)^T \gamma_\alpha^T\right]\\
&= -{\rm Tr}\left[S(x,y|U^*)\gamma_\beta S(y,\hat{\mu}|U^*) (U^*)^\dagger_\mu(0)\gamma_{\nu}
S(0,x|U^*)\gamma_\alpha\right]\ .  \nonumber
\end{align}
Using the previous equation and averaging over gauge configurations, we get
\begin{equation}
\left\langle{\rm Tr}[\gamma_{\alpha}S(x,0)
\Gamma^{\cal O}_{\mu\nu}(0,z)S(z,y)\gamma_{\beta}S(y,x)]\right\rangle_g 
=\left\langle{\rm Tr}[\gamma_{\beta}S(y,z)\Gamma^{\cal O}_{\mu\nu}(z,0)S(0,x)
\gamma_{\alpha}S(x,y)]\right\rangle_g \ ,  
\end{equation}
and similarly for $x \leftrightarrow y$ and $\alpha
\leftrightarrow \beta$. 
Finally, we arrive at the following expression used for actual
calculation:
\begin{multline}
\langle \gamma(p\lambda')|{\cal O}_{\mu\nu}(0)| \gamma(p\lambda)\rangle^{\rm(latt)}
_{\rm conn.}
= -e^2\epsilon^*_\alpha(\lambda')\epsilon_\beta(\lambda)\sum_{x,y}
         e^{\omega (x_4-y_4)} e^{-i\vec{p} \cdot (\vec{x}-\vec{y})} 
 \sum_{z,f} e_f^4  (2\kappa)^3 \times \nonumber \\
2 \cdot \left\langle{\rm Tr}[\gamma_{\alpha}S(x,0)
\Gamma^{\cal O}_{\mu\nu}(0,z)S(z,y)\gamma_{\beta}S(y,x)] 
 + {\rm Tr}[\gamma_{\beta}S(y,0)\Gamma^{\cal O}_{\mu\nu}(0,z)S(z,x)
\gamma_{\alpha}S(x,y)] \right\rangle_g .
\end{multline}

The above matrix element is evaluated by summing over the propagators
starting from point sources placed at the position of ${\cal O}$ 
instead of one of the electromagnetic vertices.
The reason for this strategy, different from hadronic structure functions
\cite{martinelli,schierholz,negele},
is that, as evident from Eq.~(\ref{eq:trace}), the electromagnetic currents 
at both source and sink should be summed over {\it all} the time slices with 
proper momentum dependence, in contrast
to the hadron structure functions where the time coordinates 
of the hadron operators are fixed. [Because of the translational
invariance in the time direction, one can still fix the time coordinate of 
one of the electromagnetic vertices instead of the operator 
observable. The answer shall not be much different if the time
direction is sufficiently large.]

In practical simulations, two sequential propagators are generated 
from each of the $12=(4\times 3)$ point sources 
\begin{eqnarray}
&M_{\alpha}(\vec{p},\omega,y) = \sum_x S(y,x)\gamma_{\alpha}S(x,0)e^{-i\vec{p}
          \cdot\vec{x}+\omega x_4} \ , \nonumber \\
&M'_{\beta}(\vec{p},\omega,x) = \sum_y S(x,y)\gamma_{\beta}S(y,0)
        e^{+i\vec{p}\cdot\vec{y}-\omega y_4} \ . 
\end{eqnarray}
The complementary propagator generated from the operator $\cal O$
is 
\[
N^{\cal O}_{\mu\nu}(x) = \sum_z \Gamma_{\mu\nu}^{\cal O}(0,z)S(z,x)\ , 
\]
which can be calculated from the point sources
as $N_{\mu\nu}^{\cal O\dagger}(x)= \sum_z \gamma_5S(x,z)\gamma_5 
\Gamma^{\cal O\dagger}_{\mu\nu}(z,0)$. 
In terms of the above propagators,
\begin{multline}
 \langle x \rangle^{\gamma}_{\rm conn.} /\alpha_{\rm em} = 
-\frac{4\pi}{2p_{\alpha} p_{\beta}}\sum_{f}e_f^4 
\epsilon_{\mu}^* \epsilon_{\nu} (2\kappa)^3\\
\times 2 \cdot \biggl[\sum_y{\rm Tr}[N^{\cal O}_{\mu\nu}(y)\gamma_{\beta}
M_{\alpha}(y)]e^{i\vec{p}\cdot\vec{y}-\omega y_4}
+\sum_x{\rm Tr}[N^{\cal O}_{\mu\nu}(x)\gamma_{\alpha}M'_{\beta}(x)]
e^{-i\vec{p}\cdot\vec{x}+\omega x_4}\biggr] \ ,
\end{multline}
where $\alpha$ and $\beta$ are fixed indices (without summation)
and the polarization vector can be chosen as a linear
one in the $x$ direction,
$\epsilon_\mu=(1,0,0,0)$. For 3 degenerate quarks, $\sum_f e_f^4$ = $(-1/3)^4
+(2/3)^4+(-1/3)^4$ = 2/9.
[Alternatively, one can generate doubly sequential propagators:
\begin{eqnarray}
M_{(2)\alpha\beta}(z) = \sum_{x,y} S(z,y)\gamma_{\beta}S(y,x)
   \gamma_{\alpha}S(x,0)e^{i\vec{p}\cdot\vec{y}-\omega y_4}
      e^{-i\vec{p}\cdot\vec{x}+\omega x_4} \ , \\
M'_{(2)\alpha\beta}(z) = \sum_{x,y} S(z,x)\gamma_{\alpha}S(x,y)
   \gamma_{\beta}S(y,0)e^{i\vec{p}\cdot\vec{y}-\omega y_4}
     e^{-i\vec{p}\cdot\vec{x}+\omega x_4} \ , \nonumber
\end{eqnarray}
which would eliminate the need for separate inversions 
for different operators. However, the successive inversions are then 
not desirable numerically when high precision is needed.]

We use $\beta=6.0, 16^3 \times 32$ quenched configurations generated by Kilcup 
et al. \cite{kilcup}, available from NERSC lattice archive. The lattice spacing 
is determined by taking the $\rho$ meson mass at the chiral limit: $a^{-1} 
\sim$ 2.4~GeV \cite{mass98}. To decide on the boundary condition
for the Wilson fermion in time direction, we made a test run with 
$\beta=5.7, 8^2\times 16 \times 32 $ lattices. For the 
antiperiodic boundary condition, the finite size effect
gives a noisy $\langle x\rangle^\gamma$ at a $m_q \sim$ 200~MeV
$(\kappa=0.16)$. On the other hand, the effect is under much better
control for a $m_q \sim 100$~MeV $(\kappa=0.165)$ for 
the Dirichlet boundary 
condition. For heavier quark masses, both boundary conditions 
give consistent results. Therefore, we use a Dirichlet boundary
condition to obtain the main result of the paper.

The photon momentum fraction $\langle x\rangle^{\gamma}$ is 
measured by evaluating $\langle \gamma(p\lambda)|{\cal O}_{34}(0)| 
\gamma(p\lambda)\rangle$
for the  momentum $\vec{k} = (0,0,2\pi/16),~\omega = 2\pi/16$.
We use hopping parameters $\kappa=$ 153, 0.154, 0.155, which corresponds to 
pion masses $m_{\pi} \sim$ 0.424, 0.364, 0.301 in lattice units, respectively \cite{mass98}.
We also evaluate $\langle x\rangle^{\gamma}$ for $\vec{k} = (0,0,\pi/16), 
~\omega = \pi/16$ ($\kappa=$ 154, 0.155), on the same Monte
Carlo lattices, which is possible by employing the antiperiodic boundary
condition for the pseudofermion fields in the $z$ direction.
 
To be close to the real world, the moment is calculated 
for three flavors of quarks with the degenerate masses. 
In the quenched approximation, the ``$\rho$ meson'' is the lowest 
hadronic state with photon quantum number.
With these parameters, $m_{\rho}$  
= 0.426 (in lattice units) for the lightest quark mass ($\kappa=0.155$)\cite{mass98} and $\sqrt{m_{\rho}^2+\vec{k}^2}$ = 0.577.
This gives $ L_t\times\left(\sqrt{m_{\rho}^2+\vec{k}^2}-\omega\right) \sim 5$.

\begin{figure}[hbt]
\begin{center}
\epsfig{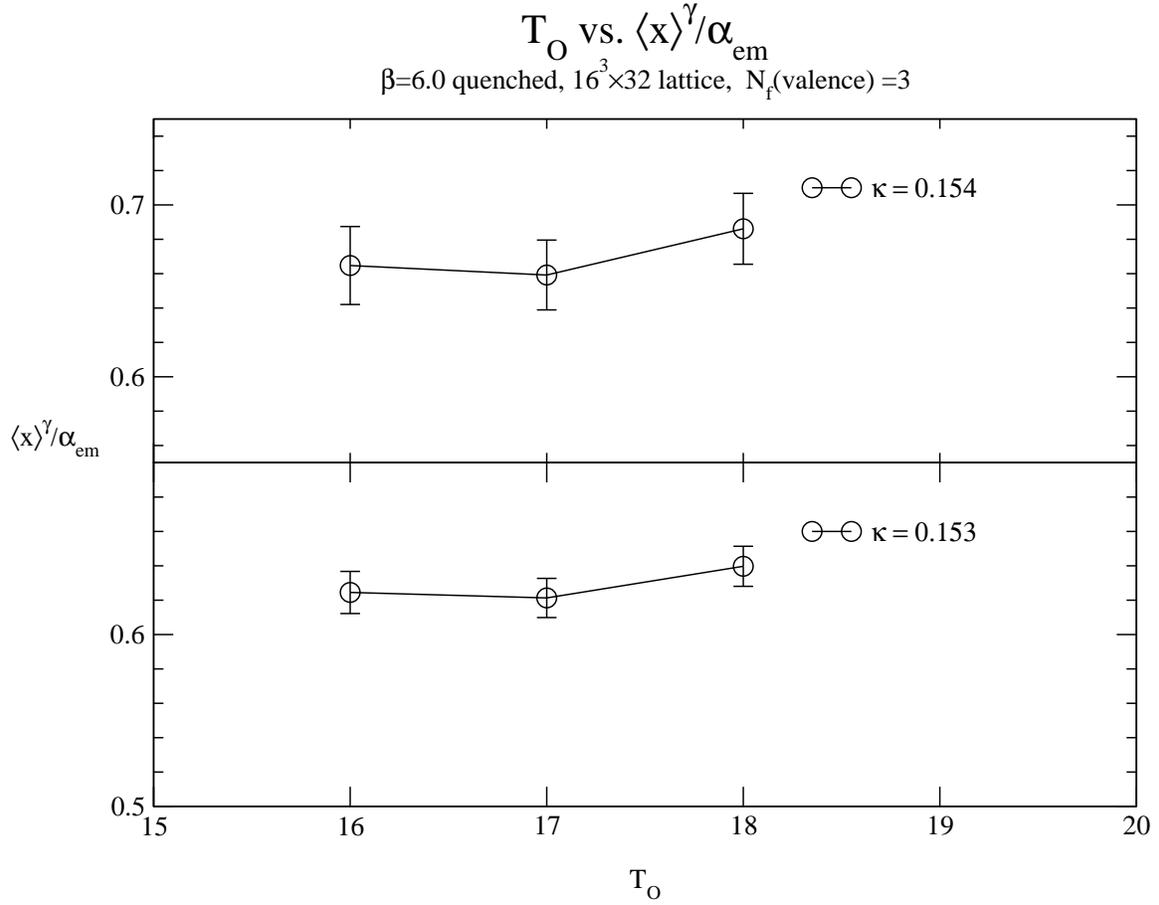}
\vspace{0.3in}
\end{center}
\caption{The first moment of $F_2^{\gamma}$ for $\beta$=6.0
quenched configurations for the different time position of the operator
($T_{\cal O}$). }
\label{fig_f1}
\end{figure}

To test if this is large enough, we measured  $\langle x \rangle^{\gamma}$
with 3 different time positions of the operator ${\cal O}$.
Figure 1 shows the value of the first moment of $F_2^{\gamma}$ for 
$\kappa=0.153,0.154$ with $T_{\cal O}$ = 16,17,18 (time coordinate runs from
0 to 31). The numerical values of $\langle x\rangle^{\gamma}$ for 
different $T_{\cal O}$'s are consistent
with each other (Fig. \ref{fig_f1}), which shows the temporal size of the box is 
large enough for the mass range. Note that the lattice result  
has been converted to a continuum renormalization scheme ($\overline{MS}$ 
in this case) by
\begin{gather}
{\cal O}^{\overline{\rm MS}}_{\mu\nu}(Q^2) = Z_{\cal O} \cdot {\cal 
O}^{\rm LATT}_{\mu\nu}(a^2)\\
Z_{\cal O}=1+\frac{g_0^2}{16\pi^2}\frac{N_c^2-1}{2N_c}
\left(\gamma^{\overline{\rm MS}} \log(Q^2a^2)-(B^{\rm LATT}-
B^{\overline{\rm MS}})\right) 
\end{gather}
Here we use the renormalization constant calculated perturbatively for $Q = a^{-1}$
\cite{capitani}, $Z_{\cal O} =  0.9892$. (The mixing between the operator 
${\cal O}_{\mu\nu}$ (Eq.~(\ref{contoper})) and gluon operator ${\rm
Tr}[F_{\mu\rho}F_{\rho\nu}]$ is absent in the quenched approximation.)

\begin{figure}[hbt]
\epsfig{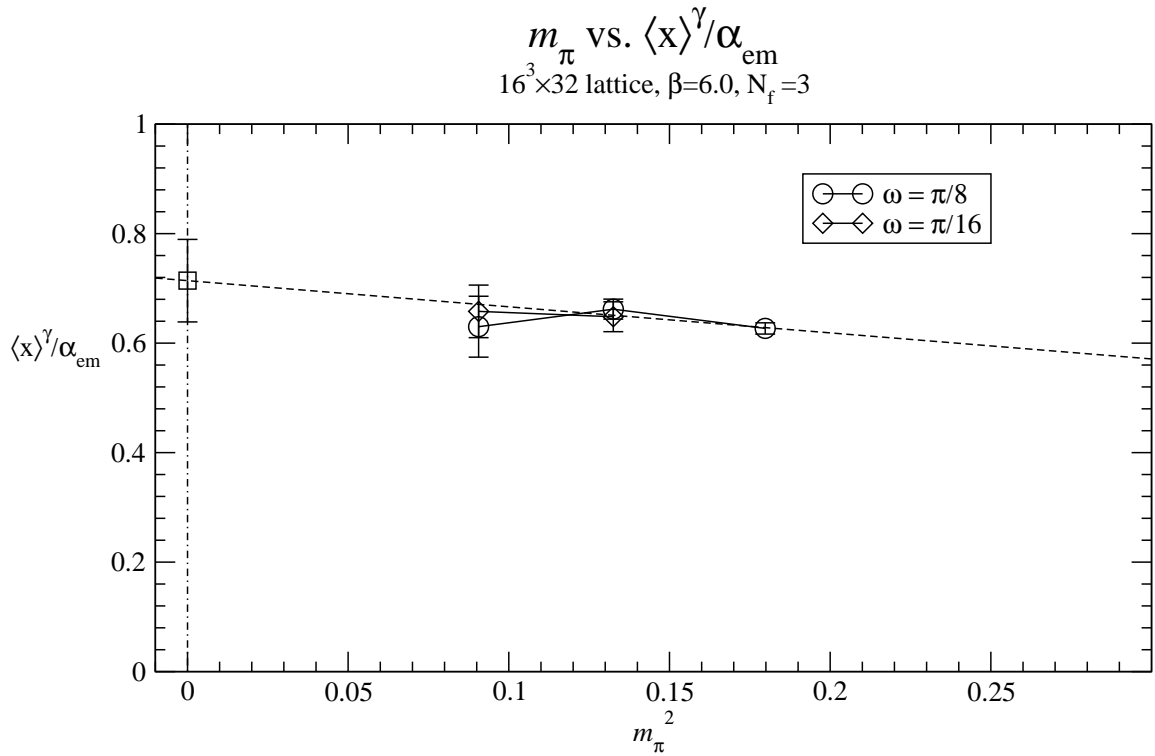}
\vspace{0.3in}
\caption{The first moment of $F_2^{\gamma}$ for $\beta$=6.0
quenched configurations $(a^{-1} \sim 2.4~$GeV). $\omega = \pi/8 \ (\pi/16)$
correspond to the periodic (antiperiodic) boundary condition  for the
pseudofermion fields in the $z$ direction. $m_\pi$ is in lattice units ($a$ = 1).}
\label{fig_f2}
\end{figure}

Figure \ref{fig_f2} shows the first moment of $F_2^{\gamma}$ for 3 degenerate 
quark flavors. The values of $\langle x \rangle^{\gamma}/\alpha_{\rm em}$ for
both momenta ($\omega = 2\pi/16, \pi/16$) are well within the error bar.
A linear fit to the chiral limit ($m_\pi=0$) 
gives $\sim$ 0.72(8), which is significantly larger than
existing experimental results \cite{nisius}.
Also, existing theoretical models such as the GRV model \cite{GRV} and the Quark Parton
Model (QPM) \cite{QPM} predicts 
$\langle x \rangle^{\gamma}/\alpha_{\rm em}$
$\sim$ (0.3$-$0.4) for the value of $Q^2$ investigated here.

We notice that
the previous lattice studies of hadron structure functions 
\cite{martinelli,schierholz,negele} showed a consistent overestimation of 
the first moment of the structure functions ($\langle x \rangle$).
It has been suspected that the absence of sea quarks enhances the valence 
quark contribution, resulting in larger values of $\langle x\rangle$.
However, recent results from unquenched structure function studies \cite{negele} 
show no significant deviation from quenched results, which suggests that  much smaller quark masses
and/or a larger lattice size, quenched or unquenched, may be necessary to approach
the continuum value. So some amount of overestimation for $\langle x\rangle^\gamma$ is not unexpected.
Still, the difference is more significant than those for hadron structure
functions. This may suggest that $F_2^\gamma$ is actually larger at the
extremes of $x$, where the experimental data is not available.

Another source of discrepancy is that the contribution
from disconnected diagrams may be 
large. We are currently investigating this. 
[It should be noted that for 3 degenerate (up, down and strange) quarks, 
the trace of electromagnetic operator, ${\rm Tr}[J_\beta(x)]$, 
is zero since $\sum_f e_f=0$; 
we would only need to evaluate the diagrams similar to 
disconnected insertion (DI) diagrams studied for quantities, such 
as the strangeness magnetic moment of the nucleon \cite{liu}.] 

In summary, the formalism necessary to compute the moments 
of the photon structure function is presented. 
We then compute the first moment of the unpolarized 
photon structure function $F_2^{\gamma}$ on quenched 
 $\beta = 6.0$ ($Q \sim a^{-1} \sim$ 2.4 GeV) lattice configurations.
The result is somewhat larger than the existing 
theoretical models and experimental results. 
The discrepancy is likely due to the quenched approximation
and the disconnected contribution.
If the result persists, it may indicate 
$F_2^{\gamma}(x)$ is significantly higher at near $x=0$ or $x=1$
than the theoretical models show. However,
the systematic errors present in the measurement need to be 
understood to make more meaningful comparison.

This work is supported by funds provided by the U.S. 
Department of Energy (DOE)
under grant No. DOE-FG02-94ER-40762.
The numerical calculation reported here
was performed on the Calico Alpha Linux Cluster 
at the Jefferson Laboratory, Virginia. 

\end{document}